# Fisica e Metafisica?
# The Science at the time of de Chirico and Carrà

Susanna Bertelli - Università degli Studi di Ferrara, Dipartimento di Fisica e Scienze della Terra – susanna.bertelli@unife.it
Paolo Lenisa - Università degli Studi di Ferrara, Dipartimento di Fisica e Scienze della Terra – paolo.lenisa@unife.it
Grazia Zini- ex Physics Professor of Università degli Studi di Ferrara, Dipartimento di Fisica e Scienze della Terra – zini@fe.infn.it

*Abstract*: The evolution of Physics between the second half of the XIX century and the beginning of the XX century is presented coupling art and science, in the framework of Ferrara. The main characters are Giuseppe Bongiovanni, Professor of Experimental Physics and Giorgio de Chirico, artist founder of the Metaphysical art movement.

## 1. Fisica e Metafisica?

### *1.1. The science filter applied to the Metaphysical art*

*Fisica e Metafisica?* is the title of a temporary scientific exhibition dedicated to the historical instruments, scientists and discoveries that led to the birth of Modern Physics. This event was held in Ferrara between 2015 and 2016 [Fisica e Metafisica]. The event was organized by the Department of Physics and Earth Sciences, the University Museum System of the University of Ferrara and the National Institute for Nuclear Physics (INFN). The instruments exposed in the exhibition belong to the *Historical Physics Instruments Collection*[1] [CISFIS] which is a section of the University Museum System cited above, that houses part of the physics instruments used by professors of Physics of the Ferrara University over two centuries.

    This event took place in conjuction with the art exhibition "*De Chirico a Ferrara, Metafisica e avanguardie*" [De Chirico a Ferrara] devoted to the Metaphysical current, that developed in Ferrara during the years of the World War I. The interaction point of these two events is the friendship between Giuseppe Bongiovanni, Professor of Experimental Physics at the University of Ferrara and Giorgio de Chirico, founder of the Metaphysical art movement.

---

[1] Collezione Instrumentaria delle Scienze Fisiche, section of University Museum System of Ferrara.

The title *Fisica and Metafisica?* refers to the presence of intriguing elements in the Metaphysical masterpieces. Some of these paintings, in fact, contain components that are clearly scientific instruments and some others show elements that, when seen by the eye of a physicist, recall devices used in a Physics laboratory. Such "science filter" allowed the authors of this article to find strong correspondence between the images and the instruments of the *CISFIS* Collection that were present in the Physics Cabinet of the University at the time of the direction of Professor Bongiovanni (Bongiovanni 1898), (Bongiovanni 1900), (Bottoni 1892), (Caracciolo, Zini 2009). He met Giorgio de Chirico and the group of artists and intellectuals that formed in Ferrara between 1915 and 1918, including Alberto Savinio (brother of G. de Chirico), Filippo de Pisis, Corrado Govoni, Giuseppe Ravegnani and Carlo Carrà. They referred to Bongiovanni as "*the astronomer*" and the Professor introduced them to the Physics he was studying, since Bongiovanni himself and some of the instruments he used for research, are mentioned in several works of the prose and poetry of Savinio[2], Giorgio de Chirico and Filippo de Pisis[3].

The "*Fisica and Metafisica?*" exhibition was therefore realized using these "presences" in the metaphysical pictures as a starting point to describe the development of Modern Physics in choosen sections: measurements and prototypes of measurements, meteorology, electromagnetism, astronomy and medical physics. Each section was matched with metaphysical paintings (authorized reproductions) and archive documents that testify the development of Physics in Ferrara.

After a brief presentation of Giuseppe Bongiovanni and Giorgio de Chirico, the sections and the related instruments of the exhibition are here described.

### *1.2. Giuseppe Bongiovanni and the Experimental Science*

Giuseppe Bongiovanni (Lugo di Romagna 1851- Siena 1918) was a scientist well known in the national and international scientific environment (Graziani 2000). The research fields he was involved in were different, including mechanics, meteorology, electromagnetism, astronomy and he was among the founders of the most important scientific societies and academies in Italy and in Europe.[4] He took the degree in Mathematics and Physical Sciences in 1873, in Pisa and between 1877 and 1884 he became Professor of Experimental Physics at the Regio Liceo of Ferrara and at the Ferrara University. He was the Director of the Physics Cabinet and the Meteorological and Seismic Observatory located in Palazzo Paradiso, the University headquarters. In 1896, the Observatory was moved to the Este Castle in the tower of Santa Caterina. Bongiovanni obtained many instruments from the Central Meteorological Office of Rome and he invented and renewed few of them, keeping the research in Physics at the leading edge. The height of the Observatory's splendor was during his direction. He wrote fourty papers ranging from mechanics to electromagnetism.

---

[2] Alberto Savinio mentions Bongiovanni in two lyrics of Hermaphrodito.(Savinio 1918).
[3] Filippo de Pisis mentions Bongiovanni in the text I Predestinati that is included in Confessioni (De Pisis 1996).
[4] Bongiovanni was member of the Medical and Natural Sciences Academy of Ferrara, the French Astronomical Society and the Italian Astronomical Society, he was one of the founders of the Italian Physical Society and of the Italian Seismological Society.

### *1.2. Giorgio de Chirico in Ferrara, the "Metaphysical town"*

Giorgio de Chirico (Volos, Greece 1888-Rome 1978) was the founder of the metaphysical art. In 1915, de Chirico and his brother, Savinio, were enlisted into the Italian army to fight in World War I and stationed at Ferrara. In this city he found himself *"assailed by revelations and inspirations"*. Due to nervous disorders, Giorgio de Chirico [Fondazione de Chirico] and Carlo Carrà were admitted into a military hospital *Villa del seminario,* where they continued painting thanks to the Director, Gaetano Boschi (Boschi 1918), who assigned the artists a room as a studio, where they could work and through their exchanges Metaphysical art, *pittura metafisica*, was born. The bond Art-and-Science is employed by Giorgio de Chirico when he defines the Metaphysical abstraction using the X-rays claiming that:

> "Deducendo si può concludere che ogni cosa abbia due aspetti: uno quello corrente che vediamo quasi sempre, e che vedono gli uomini in generale, l'altro lo spettrale o metafisico che non possono vedere che rari individui in certi momenti di chiaroveggenza e di astrazione metafisica, così come certi corpi occultati da materia impenetrabile ai raggi solari non possono apparire che sotto la potenza di luci artificiali quali sarebbero i raggi X, per esempio" (De Chirico 1919).

### 2. The sections of the exhibition

In this part the areas of the *Fisica e Metafisica?* exhibition are illustrated, presenting the metaphysical paintings matched to the sections and the instruments showed.

### *2.1 Measurements and prototypes of measurements*

This section is introduced by a painting of Carlo Carrà named *The Enchanted Chamber* (1917). In this painting there is a small weight and there is a picture that recalls a sextant, an instrument to measure the angle between two objects. Leading from this painting, a section of the exhibition has been dedicated to the definition of measurement, the units of measurements and a brief timeline on the history of measurements, starting from the anthropomorphic units to the International System of Units. This section housed prototypes of the litre, a *libbra metrica campione*, a metre, an half-metre converter ruler. The half-metre converter ruler was made by Pietro Torquato Tasso, an artisan of Ferrara. This instrument testifies the introduction of the metric system in Italy (1811). In each of its faces there are different systems of unit: *piede di Parigi, piede di Ferrara, due palmi romani, quarto di canna di mercante romana, quarto d'auna di Parigi, quarto d'auna di Londra,* according to the large variety of system of length units used (even in the same city) before the metric system was implemented.

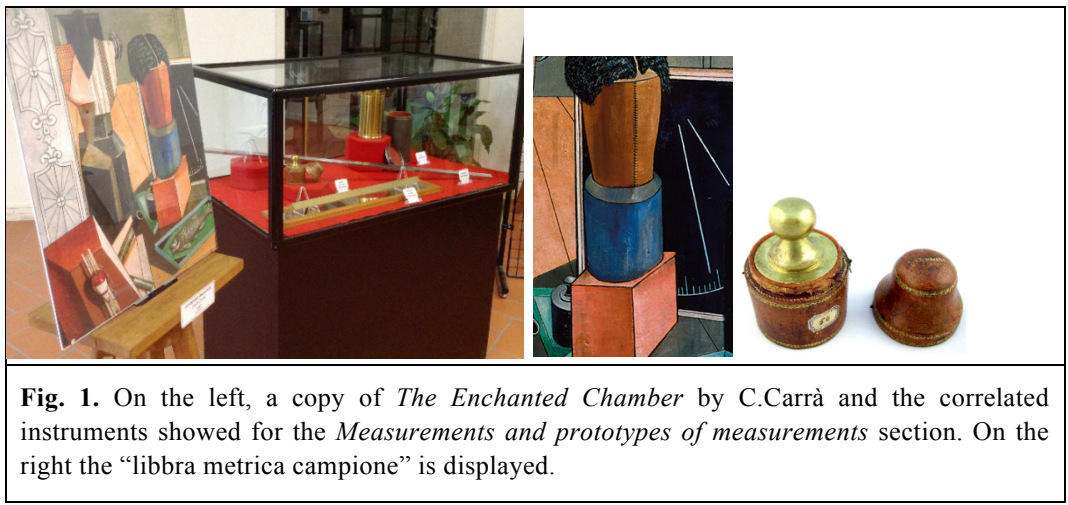

**Fig. 1.** On the left, a copy of *The Enchanted Chamber* by C.Carrà and the correlated instruments showed for the *Measurements and prototypes of measurements* section. On the right the "libbra metrica campione" is displayed.

*2.2 Meteorology*

The section Meteorology is coupled with two pictures by Giorgio de Chirico: *The Amusements of a Young Girl, 1915* and The dream of Tobias (Fig.2), 1917. In the first one, de Chirico painted a detail of the Este Castle where the tower of Santa Caterina is present. This tower was the location of the Observatory directed by Bongiovanni. In the second painting, one can notice a thermometer represented in the middle. Starting from these elements, the evolution of Meteorology in Europe and in Italy during the XIX century is described through the instruments that Bongiovanni and his predecessor, Curzio Buzzetti (1815-1877), used to equip the Observatory (Caracciolo, Zini 2009). The instruments showed during the exhibition were: an *eliofanometro*, a mercury thermometer and a maximum/minimum thermometer, a Fortin mercury barometer, condensation hygrometers[5] (by F.lli Brassart [6] and by L.Golaz) and a psychrometer[7]. Bongiovanni realized a study of the climate of Ferrara, based on several years of measurements published in 1900 (Bongiovanni 1900) and followed by annual reports up to 1911. When he became the Director of the Observatory, Ferrara already had a great tradition in the field of Meteorology. He continued the research of Curzio.Buzzetti inheriting the instruments existing in the Physics Cabinet. He optimized some of them and he also construct new ones obtaining several awards for these works. He recorded data four times a day (Bongiovanni 1900, pg. 13) and to send them periodically to the Office of Meteorology in Rome. He organized a network of eight observatories in the territory of Ferrara. Bongiovanni describes the use of the eliofanometro (Bongiovanni, 1900 pp. 79-82) to measure the hours of sun during a day. This device acts as a lens that converges the impinging sun light in a spot that leaves a burned track on a designed paper:

> "Insolazione – 1. Instrumento, esposizione e modo di osservazione
> La misura del tempo durante cui il sole ogni giorno resta scoperto dalle nubi è stata cominciata nel 1889 con un eliofanometro avuto in dono dal R. Ufficio centrale di Meteorologia …..Il cartoncino non comincia ad essere bruciato la mattina che quando il sole ha raggiunto una certa all'altezza sull'orizzonte, e la sera cessa di essere bruciato prima che il sole arrivi all'orizzonte, in causa del grande assorbimento della radiazione solare per opera dei vapori atmosferici, della nebbia e della caligine. (Bongiovanni 1900, pp 79-82)."

---

[5] Used to measure the relative humidity of the air.

[6] The Brassarts designed and built several instruments for the Central Office of Meteorology in Rome [Osservatorio Astronomico Palermo].

[7] Device that consists in two thermometers, one wet and the other dry, to measure humidity.

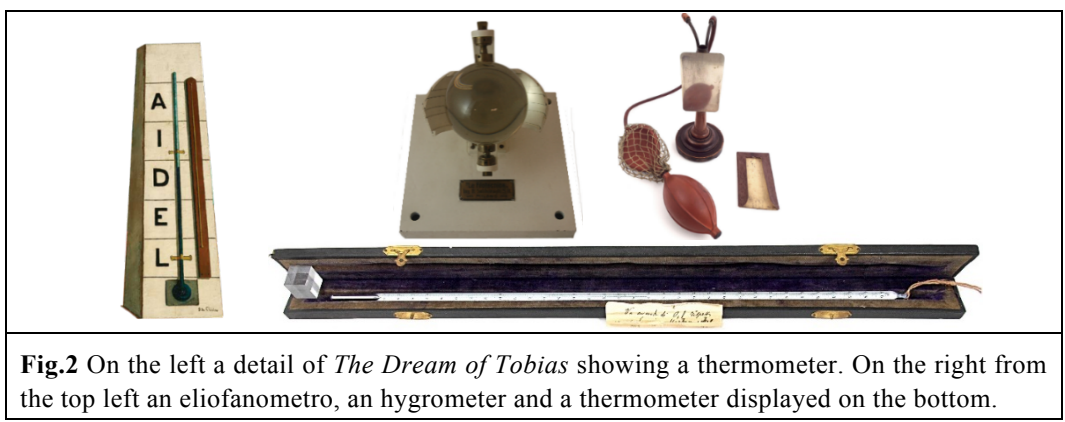

**Fig.2** On the left a detail of *The Dream of Tobias* showing a thermometer. On the right from the top left an eliofanometro, an hygrometer and a thermometer displayed on the bottom.

### 2.3. Electromagnetism

*The Jewish Angel* (1916) by Giorgio de Chirico is the painting that introduces the section devoted to electromagnetism. The connection between metaphysical art and science is given by an instrument in the lower part of the painting that recall a voltmeter (Fig.3). The progress in electricity and magnetism from the first electrostatic machines to the main applications of electromagnetism are described through these instruments of the collection: a Wimshurst machine, an electric egg, a Ruhmkorff coil, a spherical capacitor, a voltmeter, a Melloni apparatus, a model of a thermopile, a Marconi radio telegraph. The spherical capacitor was used by Bongiovanni to study the electrical insulators as described in one of his articles (Bongiovanni, 1898-99). The Marconi radio telegraph was used by Bongiovanni to send the recorded meteorogical data to Rome. Bongiovanni dedicated several papers to electromagnetism, such as *L'elettricità e la teoria elettromagnetica della luce* (1890), *Elettrologia* (1893), *Magnetismo* (1895), *I progressi della telegrafia senza fili* (1904). The Melloni apparatus was used to study the properties of the thermal radiation (calore raggiante). The equipement includes many items and also a thermopile, device used to convert thermal energy into electrical energy. The wood model of the thermopile is signed by Bongiovanni.

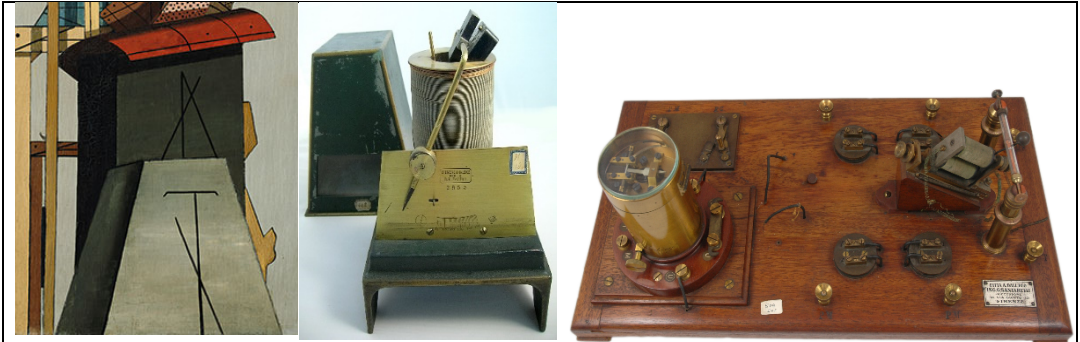

**Fig.5** On the left a detail of the painting *The Jewish Angel* and on the voltmeter matched to this painting. On the right the Marconi radio telegraph.

## *2.4 Astronomy*

The section dedicated to Astronomy is introduced by the paintings *The Philosopher and the poet* (1916). In this painting within the painting, celestial bodies are represented. In this section a timeline of Astronomy is depicted through these instruments: drawing compasses, lenses for telescope, a star pointer, a *cannocchiale*, a tellurium, (by S.Zavaglia, 1855). The telescope used by G.Bongiovanni is no longer present in the CISFIS Collection, just the lenses[8] are still present. Giorgio de Chirico refers to this telescope the poetry "La Notte Misteriosa" (1916) dedicated to the *astronomer* Bongiovanni. The tellurium shows the relative motions of the Earth, Sun and Moon and it is used to explain astronomical phenomena like alternation of day and night, the changes of the seasons, the lunar phases and the eclipses. This model was made by Sebastiano Zavaglia (Molinella 1824-Bologna 1876) and he entitled it "*Motion of the Earth and the Moon around the Sun according to the system of Copernicus, 1855, N 1*.

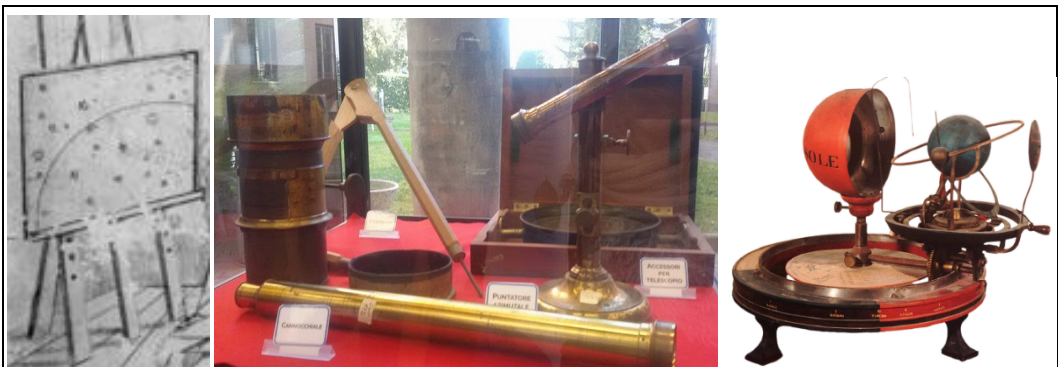

**Fig. 5.** On the left a detail of *The Philosopher and the poet* by G.de Chirico, in the middle the instruments exposed, and on the right the tellurium is displayed.

---

[8] One element is a Petzval lens, (see fig.5, the first object from the left placed vertically).

*2.4 Medical Physics*

The painting *Mother and Son* (1917) by Carlo Carrà is matched with this section in which a timeline of Medical Physics is described from the application of electricity to health care to the discovery of X rays and medicine. The connection to the metaphysical art is given by a spool in the painting that recalls a Ruhmkorff coil used to power the X rays tubes. The instruments showed in this section are a Clarke's machine, Matteucci's spiral plates, a Ruhmkorff coil, a X-rays tube, a Crookes tube, a cryptoscope.

The Clarke's machine is one of example of using electricity for health care to cure nervous diseases and for electrotheraphy, whereas the Matteucci's spiral plates were used for the magnetotheraphy. The discovery of X-rays by Roentgen is illustrated by showing the X-rays tubes of the Physics Cabinet of Ferrara. As reported by a journal of 1920[9] in an article by prof. Brunè referring to G.Bongiovanni:

> Così ha riprodotto le interessanti esperienze sui raggi X pochi giorni dopo l'annun
> cio della scoperta e giovandosi di alcuni tubi Crookes esistenti in gabinetto, soltanto
> un mese dopo la scoperta, ha potuto ottenere una prova radiografica…(Brunè 1920)

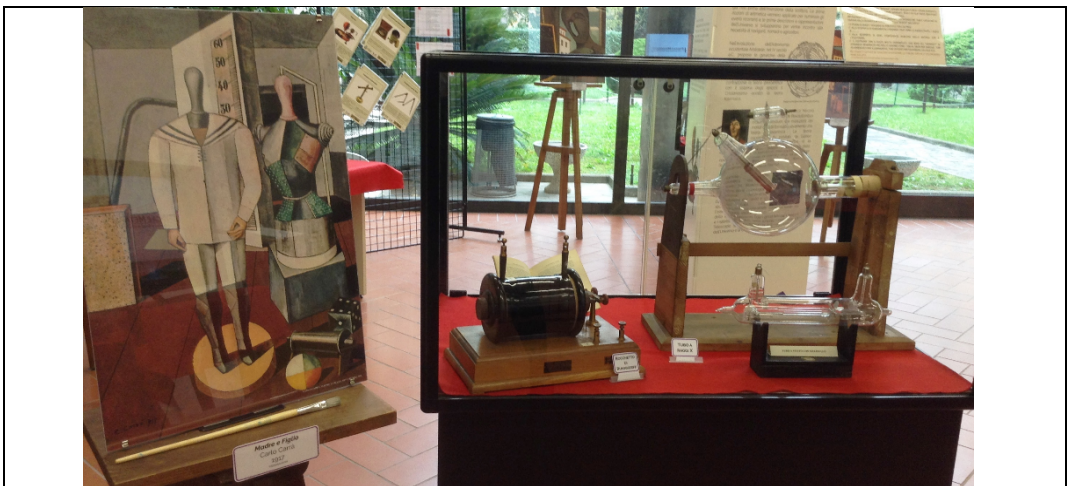

**Fig. 7.** In this picture the matching of Carlo Carrà picture to the instruments related to X rays: a Ruhmkorff coil (close to the detail of the picture that recalls it) and the X-rays tube.

**3. Conclusion**

The scientific exhibition "*Fisica e Metafisica?*" was an opportunity to display hystorical Instruments present in the Physics Cabinet of the University of Ferrara at the beginning of XX century and to chronicle the main discoveries that led to the birth of Modern Physics, coupling Art-and-Science and the history of Ferrara.

---

[9] *Gazzetta Ferrarese, 1920 The commemoration of prof. Bongiovanni*, speech by prof. Brunè.